\documentclass[twocolumn,iop,apj]{aastex63}
\usepackage{amsfonts,amsmath,graphicx,natbib,url,hyperref,nicefrac}
\usepackage{amssymb, verbatim, subfigure, paralist, soul, comment}

\hypersetup{colorlinks=true, urlcolor=blue, citecolor=blue}

\usepackage{color}

\newcommand{\ctB}{\citetalias{Bromberg+2011ApJ} }

\defcitealias{Bromberg+2011ApJ}{B11}

\shorttitle{Relativistic Jet Dynamics} 
\shortauthors{Mandal, Duffell \& Li}

\begin{document}

\title{Numerical Investigation of Dynamical and Morphological Trends in Relativistic Jets}

\author[0000-0001-9484-1262]{Soham Mandal}
\affiliation{Department of Physics and Astronomy, Purdue University, 525 Northwestern Avenue, West Lafayette, IN 47907, USA}

\author[0000-0001-7626-9629]{Paul C. Duffell}
\affiliation{Department of Physics and Astronomy, Purdue University, 525 Northwestern Avenue, West Lafayette, IN 47907, USA}

\author[0000-0001-5262-6150]{Yuan Li}
\affiliation{Department of Physics, University of North Texas, Denton, TX 76203, USA}

\begin{abstract}

Active galactic nuclei (AGN) show a range of morphologies and dynamical properties, which are determined not only by parameters intrinsic to the central engine but also their interaction with the surrounding environment. We investigate the connection of kiloparsec scale AGN jet properties to their intrinsic parameters and surroundings. This is done using a suite of 40 relativistic hydrodynamic simulations spanning a wide range of engine luminosities and opening angles. We explore AGN jet propagation with different ambient density profiles, including $r^{-2}$ (self-similar solution) and $r^{-1}$, which is more relevant for AGN host environments. While confirmation awaits future 3D studies, the Fanaroff-Riley (FR) morphological dichotomy arises naturally in our 2D models. Jets with low energy density compared to the ambient medium produce a center-brightened emissivity distribution, while emissivity from relatively higher energy density jets is dominated by the jet head. We observe recollimation shocks in our simulations that can generate bright spots along the spine of the jet, providing a possible explanation for ``knots'' observed in AGN jets. We additionally find a scaling relation between the number of knots and the jet-head-to-surroundings energy density ratio. This scaling relation is generally consistent with the observations of the jets in M87 and Cygnus A. Our model also correctly predicts M87 as FR I and Cygnus A as FR II. Our model can be used to relate jet dynamical parameters such as jet head velocity, jet opening angle, and external pressure to jet power and ambient density estimates.

\end{abstract}


\section{Introduction} \label{sec:intro}

Radio-loud AGN drive relativistic jets into their surrounding medium. These jets expand and can form structures extending up to hundreds of kiloparsecs. The emission from AGN jets have been observed over length and energy scales spanning several orders of magnitude, enabling estimates of dynamical parameters such as jet velocities \citep{Kellermann+2004ApJ,Kellermann+2007Ap&SS,Lister+2009AJ,Lister+2016AJ,Lister+2019ApJ,Jorstad+2005AJ,Jorstad+2017ApJ,Angioni+2019A&A}, opening angles \citep{Pushkarev+2009A&A,Pushkarev+2017MNRAS,Algaba+2017ApJ}, and jet power \citep{Godfrey+2013ApJ,Foschini+2019hepr}. The kiloparsec scale morphology of AGN jets is seen to have two distinct categories, despite the wide range of dynamical parameters: the center-brightened or FRI jets, and the edge-brightened or FRII jets \citep{Fanaroff_Riley1974MNRAS}.

The basic morphology of relativistic astrophysical jets has been established by several analytical \citep{Blandford+1974MNRAS,Scheuer1974MNRAS,Begelman+1989ApJ,Falle1991MNRAS,Kaiser+1997MNRAS,Matzner2003MNRAS,Lazzati+2005ApJ} and numerical studies \citep{Duncan+1994ApJ,Marti+1994A&A,Marti+1997ApJ,Komissarov+1998MNRAS}. The accepted picture is that the advancing jet forms an overpressured double bow-shock structure at its head. Material that enters the jet head is pushed out in a direction transverse to that of jet propagation, and forms a hot cocoon around the jet. This cocoon may pressurize the jet and influence its dynamics and state of collimation. The dynamics and state of collimation are the main aspects of the differences between FRI and FRII jets. The origin of this dichotomy has been studied extensively using numerical computations \citep[e.g.,][]{Rossi+2008A&A,Krause+2012MNRAS,Li+2018ApJ}.

\begin{figure}
\centering
\gridline{ \fig{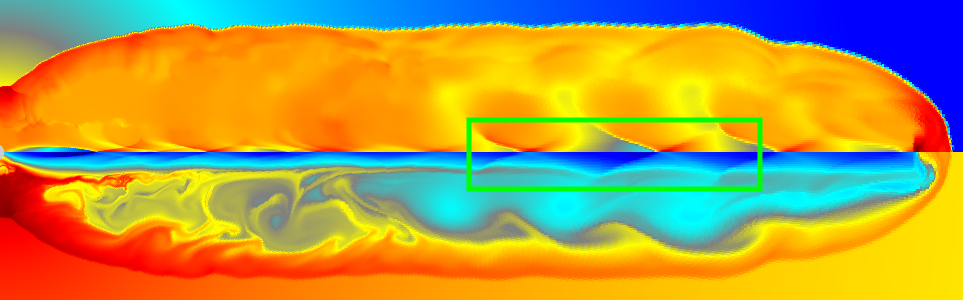}{0.45\textwidth}{} }
\gridline{ \fig{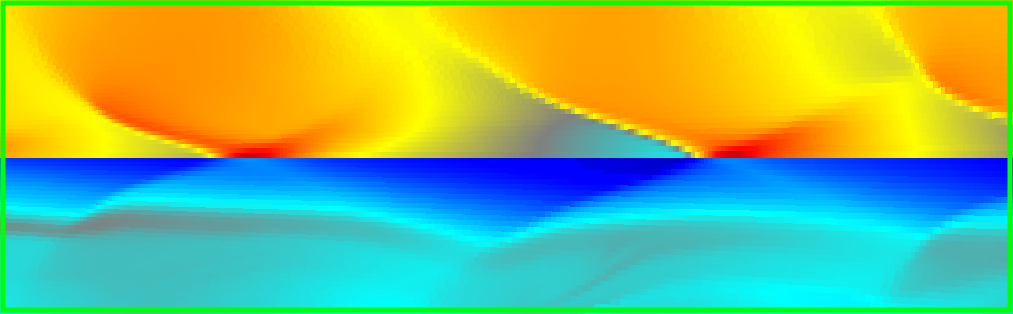}{0.45\textwidth}{} }
\caption{Snapshots showing an example of our self-similar solutions. \textit{(Top)} Logarithm of the pressure (upper half) and logarithm of the density (lower half) for a jet with $\lambda = 2.5\times 10^{-4}$ and opening angle $\theta_0 = 0.20$, expanding through an ambient medium with density profile $\rho \propto r^{-2}$. It is clearly seen that the jet experiences many recollimation shocks. \textit{(Bottom)} Zoomed-in plot for the two recollimation shocks enclosed in the green rectangle in the top panel.
\label{fig:example}}
\end{figure}

\cite{Bromberg+2011ApJ} \citepalias[hereafter][]{Bromberg+2011ApJ} developed a generalized analytical framework that characterizes non-magnetized jets expanding against a static ambient medium for widely varying dynamical (e.g., non-relativistic vs relativistic jet head velocities) and morphological (e.g., collimated vs freely expanding jets) properties. Their model provides solutions for important dynamical parameters of the jet such as the head velocity, cocoon pressure etc. in terms of power laws. It also provides the criteria that determine the specific nature of the jet dynamics and morphology, which in turn determine the exact form of the power laws. The power law solutions provided by \ctB have been verified numerically by \cite{Mizuta+2013ApJ} and \cite{Harrison+2018MNRAS}. The latter also calibrate the numerical coefficients for the power laws and discuss the applicability of 2D axisymmetric simulations for deriving them.

The aim of this work is to use the model developed by \ctB along with our own numerical calculations to infer dynamical parameters of AGN jets from observational features seen across the known AGN population. We proceed by first testing \ctB's power law scalings in the context of a relativistic jet for a wide range of jet luminosities and initial opening angles. This is done by numerically calculating self-similar solutions for the system. We then use our findings to obtain quantities and/or morphological properties that can be directly related to observations. We also show that our results can be applied to realistic AGN jets, which aren't necessarily self-similar.


We briefly introduce the jet model and scalings provided in \ctB in Section~\ref{sec:anly}. The numerical method and parameter space adopted in this work are described in Section~\ref{sec:numerical}. The methods of diagnosis we use to analyze our numerical solutions are discussed in Section~\ref{sec:diagnostics}. The results of the numerical calculations and comparison with analytical scalings are presented in Section~\ref{sec:results}. Comparison of our results to observations as well as previous studies and scope for future work are discussed in Section~\ref{sec:discussion}. We provide a brief summary in Section~\ref{sec:conclusion}.

\renewcommand{\arraystretch}{1.5}
\begin{deluxetable*}{cccc}
\tablewidth{0pt}  
\tablecaption{System characteristics in different dynamical regimes} 
\tablehead{ \colhead{Parameter} & \multicolumn{2}{c}{ Collimated Jet } & \colhead{Uncollimated Jet} }
\startdata
& Non-relativistic Head & Relativistic Head & Causal Head \\
\hline
& $\lambda<\theta_0^2$ & $\theta_0^2\ll\lambda<\theta_0^{-4/3}$ & $\theta_0^{-4/3}<\lambda\ll8\theta_0^{-4}$ \\
\hline
$\beta_h$ & $\lambda^{1/3}\theta_0^{-2/3}$ & 1 & 1 \\
$\Gamma_h$ & 1 & $\lambda^{1/5}\theta_0^{-2/5}$ & $\lambda^{1/4}$ \\
$\theta_j$ & $\lambda^{1/6}\theta_0^{5/3}$ & $\lambda^{3/10}\theta_0^{7/5}$ & $\theta_0$ \\
$\beta_c$ & $\lambda^{1/3}\theta_0^{1/3}$ & $\lambda^{1/5}\theta_0^{3/5}$ & $\lambda^{1/8}\theta_0^{1/2}$ \\
$\mathrm{P_c}/\rho_a$ & $\lambda^{2/3}\theta_0^{2/3}$ & $\lambda^{2/5}\theta_0^{6/5}$ & $\lambda^{1/4}\theta_0$ \\
$\Tilde{L}$ & $\lambda^{2/3}\theta_0^{-4/3}$ & $\lambda^{2/5}\theta_0^{-4/5}$ & $\lambda$ \\
\enddata
\label{t:lambdascaling}
\tablecomments{Reformulation of the scalings from first three columns in Table 1 of \cite{Bromberg+2011ApJ} in terms of $\lambda$ and $\theta_0$. We use our simulations to verify scalings for the following quantities:  the jet head speed ($\beta_h$) and Lorentz factor ($\Gamma_h$), the cocoon expansion speed ($\beta_c$), opening angle of the jet ($\theta_j$), and the ratio of cocoon pressure to ambient density at the jet head ($\mathrm{P_c}/\rho_a$). Note that $\Gamma_j$ denotes the intrinsic Lorentz factor of the jet and in general differs from $\Gamma_h$. Scalings for {$\Tilde{L}$} are also provided for ease of comparison. }
\end{deluxetable*}

\section{Analytical considerations} \label{sec:anly}

A brief qualitative overview of the jet model proposed by \ctB is presented here. The reader is referred to the original work for a detailed explanation, as well as to \cite{Harrison+2018MNRAS} for an excellent quantitative overview. The jet head moves into the ambient medium with speed $\beta_h$. As mentioned before, the jet head expels material to generate a hot pressured cocoon around the jet. The cocoon expands laterally with speed $\beta_c$. It drives a reverse shock into the jet towards its axis that, subject to system parameters, may cause collimation of the jet. The opening angle of the jet is denoted by $\theta_j$, which is generally different from the engine opening angle $\theta_0$ (unless the jet expands freely). The cocoon is expected to have approximately uniform pressure $P_c$ when the jet is collimated, or expands preserving causal contact throughout the jet head. \ctB found that the dynamics of this system can be described in terms of two quantities, the engine opening angle $\theta_0$ and the relativistic jet-to-environment energy density ratio at the jet head ($\Tilde{L}$), given the form of the ambient density profile. They also show that the criteria that determine the dynamical behavior of the jet, for example, whether the jet stays collimated, or whether the jet head maintains an overall causal contact, can be expressed solely by relations between $\Tilde{L}$ and $\theta_0$ (with omission of numerical coefficients of order unity). Once the specific dynamical behavior of the jet is known, the dynamical parameters of the jet (i.e., $\beta_h$, $\beta_c$, $\theta_j$ etc.) also scale as power laws of $\Tilde{L}$ and $\theta_0$. The quantity $\Tilde{L}$ is defined as:

\begin{equation}
\label{eq:Ltilde}
    \Tilde{L} \equiv \frac{\rho_j h_j \Gamma_j^2}{\rho_a} \approx \frac{L_j}{\Sigma_j \rho_a(z_h) c^3},
\end{equation}

where $\rho_j$, $h_j$, $\Gamma_j$, $L_j$, $\Sigma_j$ are the jet mass density, specific enthalpy, Lorentz factor, one-sided jet luminosity, and head cross-section area respectively. The ambient mass density at the location of the jet head $z_h$ is given by $\rho_a(z_h)$. 

The different regimes of evolution of the jet are dictated by relations between $\Tilde{L}$ and $\theta_0$. For example, the condition (correct up to a factor of order of unity) that the jet remains collimated is given by:

\begin{equation}
\label{eq:coll_con}
    \Tilde{L}\lesssim\theta_0^{-4/3},
\end{equation}

We express $\Tilde{L}$ in terms of another dimensionless quantity $\lambda$ \citep{Duffell+2020ApJ} defined as follows:

\begin{equation}
\label{eq:lambda}
    \lambda \equiv \frac{L_j}{\rho z_h^2 \theta_0^2 c^3}
\end{equation}

The relation between $\lambda$ and $\Tilde{L}$ is as follows:

\begin{equation}
\label{eq:lambda_ltilde}
    \Tilde{L} = \lambda \frac{\theta_0^2}{\pi \theta_j^2}
\end{equation}

For our study we will not control $\theta_j$ but $\theta_0$, the injection angle. Therefore we will describe our results in terms of the parameter $\lambda$ instead of $\Tilde{L}$. A higher value of $\lambda$ thus implies a higher jet power, provided the ambient density profile and injection angle are kept constant. \ctB predict scalings for the jet-cocoon system in a number of different dynamical regimes, the first three of which are listed in Table~\ref{t:lambdascaling}. These regimes can be thought to differ from each other in the velocities of the jet head. We use our self-similar models to test the scalings in these regimes.

\section{Numerical setup} \label{sec:numerical}

We numerically evolve our system in time in accordance with equations of relativistic hydrodynamics in spherical coordinates:

\begin{equation}
\label{eq:SRHD1}
    \partial_{\mu} ( \rho u^{\mu} ) = S_D
\end{equation}

\begin{equation}
\label{eq:SRHD2}
    \partial_{\mu} ( \rho h u^{\mu} u^{\nu} + P g^{\mu \nu} ) = S^{\nu}
\end{equation}
where $\rho$ is proper density, $P$ is pressure, $u^{\nu}$ is the four-velocity, and $h = 1 + 4P/\rho$ is the specific enthalpy (assuming an adiabatic index, $\Hat{\gamma}=4/3$). We assume our system to be axisymmetric. These equations are scale-free in the sense that there is no length or time scale present. The source terms $S_D$ and $S^{\nu}$ model mass, energy and momentum injection by the engine. All quantities are expressed in normalized or code units, with c = 1. We also evolve a passive scalar field X using:

\begin{equation}
\label{eq:passive_scalar}
    \partial_{\mu} ( \rho X u^{\mu} ) = 0
\end{equation}

X advects with the jet and is used to distinguish the material ejected by the engine from the ambient material. We set X=1 for the jet and X=0 for the ambient medium.

The numerical evolution is performed using JET \citep{Duffell+2011ApJS,Duffell+2013ApJ}, a moving-mesh code particularly suitable for modeling relativistic radial outflows over length scales of many orders of magnitude. The moving mesh technique essentially makes the calculation Lagrangian in the radial direction, with highest resolution near the poles. This resolution is necessary to accurately capture high Lorentz factors. The grid size is chosen to have a spatial resolution given by $\Delta\theta = \Delta r/r \sim 0.001$ (The highly resolved flow is apparent in Figure \ref{fig:example}). Only a couple orders of magnitude are resolved in the radial direction at any given time, but the system is evolved over length scales of many orders of magnitude by suitably moving the grid boundaries during calculation.

\subsection{Initial conditions} \label{subsec:IC}

The ambient density profile is modeled by a power-law:

\begin{equation}
    \rho(r)=\rho_0\left(\frac{r}{r_0}\right)^{-k},
\end{equation}

where $\rho_0r_0^k$ has been set equal to 1 for all our calculations.

It is known \citep{Komissarov+1997MNRAS, Bromberg+2011ApJ, Marti2019Galax} that a constant-power nonmagnetized relativistic outflow maintains a constant speed of advance for an ambient density profile of $\rho(r)\propto r^{-2}$, while the outflow accelerates or decelerates if the density profile is steeper or shallower, respectively. Accelerating or decelerating solutions are not fully self-similar because they transition between relativistic and non-relativistic regimes. Thus we choose $k=2$ for our self-similar calculations.

Observational studies \citep{Arnaud+1984MNRAS, Blundell+1999AJ, Russell+2015MNRAS} suggest that host environments of AGN have a shallower density profile, corresponding to $k\sim1\mbox{--}1.5$. Hence we also obtain solutions for $k = 1$ and compare them with the self-similar case to investigate how applicable the self-similar solutions are in the context of more realistic interstellar environments.

We perform calculations with k=2 for a number of jet luminosity and injection angle combinations to examine the dependence of our results on $\lambda$ and $\theta_0$. A single combination of jet luminosity and injection angle is chosen to simulate a jet expanding against an ambient medium where k=1. Equation \ref{eq:lambda} implies that $\lambda$ should effectively vary inversely with the jet propagation distance for k=1 provided the jet luminosity and injection angle remain constant. Therefore the effective value of $\lambda$ in this case ($\lambda_{\mathrm{eff}}$) decreases with time as the jet's size increases. This allows us to study the behavior of the jet for a continuous range of $\lambda_{\mathrm{eff}}$.

\subsection{Engine model}

We expect the true engine to operate at scales unresolved in our calculations and be governed by poorly understood physics. Injection is thus simulated at resolved scales using a parameterized model. The source terms in equations \ref{eq:SRHD1} and \ref{eq:SRHD2} completely specify the engine. The engine is turned on at the start of computation and stays on. It is parameterized by a power ($L_0$), injection angle ($\theta_0$), injection radius ($r_0$), Lorentz factor ($\gamma_0$) and baryon loading ($\eta_0$). 

The injection radius is set to twice the inner boundary radius of the domain. We have tested the effect of varying this injection radius and find it does not affect our results so long as it is sufficiently larger than the inner boundary radius. We choose a highly relativistic engine by setting the injection Lorentz factor and the injected mass-to-energy ratio to 50 and 0.001 respectively for all of our runs. This is to model a very clean engine where nearly all baryon contamination occurs via interaction with the surrounding medium. The range of $\lambda$ values attained by the jets for a given density profile power law and injection angle are listed in Table~\ref{t:engine}. It can also be seen from the table that the range of $\lambda$ is comparable for both ambient density profiles, i.e., k=2 and k=1.

\begin{deluxetable}{ccc}
\tabletypesize{\footnotesize}
\tablewidth{0pt}  
\tablecaption{\label{t:engine} Engine parameters} 
\tablehead{ \colhead{\hspace{.75cm}k}\hspace{.75cm} &\colhead{\hspace{.3cm}$\theta_0$}\hspace{.3cm} & \colhead{$\mathrm{log}_{10}\,\lambda$}\hspace{.5cm} }
\startdata
2 & 0.10 & -3.00, -2.00, -1.00, 0.00, 1.00, 2.00     \\
\hline
2 & 0.14 & -3.29, -2.29, -1.29, -0.29, 0.71, 1.71    \\
\hline
2 & 0.20 & -5.60, -5.12, -4.60, -4.12, -3.60, -3.12, \\
  &      & -2.60, -2.12, -1.60, -1.12, -0.60, -0.12, \\
  &      & 0.40, 0.88, 1.40, 1.88, 2.40              \\
\hline
2 & 0.28 & -4.89, -4.42, -3.89, -3.42, -2.89, -1.89, \\
  &      & -0.89, 0.11, 1.11, 2.11, 3.11             \\
\hline
1 & 0.20 & -4.0$\mbox{--}$3.2\tablenotemark{a}    \\
\enddata
\tablenotetext{a}{This denotes a range of effective $\lambda$ values that the jet attains during the span of the calculation.}
\tablecomments{Simulation parameters: (1) ambient density profile power law index (k); (2) jet initial opening angle in radians ($\theta_0$); (3) common logarithm of $\lambda$.}
\end{deluxetable}

The engine is represented using a nozzle function, $g(r, \theta)$, defined as follows:

\begin{equation}
    g(r, \theta) \equiv (r/r_0)e^{-(r/r_0)^2}e^{(|cos\,\theta|-1)/\theta_0^2}/N_0
\end{equation}

where $N_0$ is the normalization factor:

\begin{equation}
    N_0 = 4\pi r_0^3 \left(1-e^{-2/\theta_0^2}\right)\theta_0^2
\end{equation}

The source terms in equations \ref{eq:SRHD1} and \ref{eq:SRHD2} are expressed in terms of the nozzle function:

\begin{equation}
    S^0 = L_0 g(r, \theta)    
\end{equation}

\begin{equation}
    S^r = S_0\sqrt{1-\gamma_0^{-2}}   
\end{equation}

\begin{equation}
    S_D = S_0/\eta_0 
\end{equation}

\begin{figure*}
\gridline{\fig{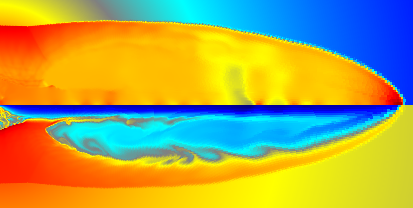}{0.32\textwidth}{(a) $\lambda=2.5\times10^{-6}$}
          \fig{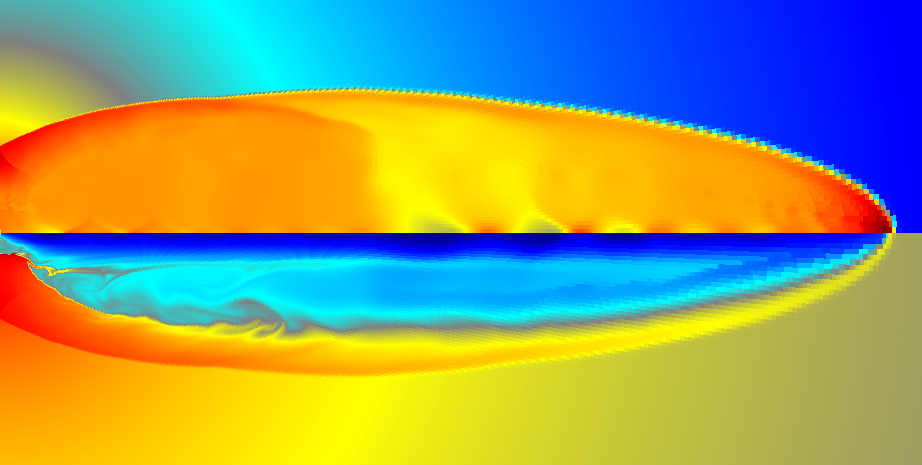}{0.32\textwidth}{(b) $\lambda=2.5\times10^{-5}$}
          \fig{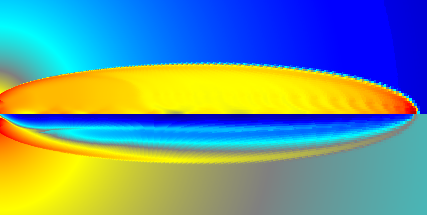}{0.32\textwidth}{(c) $\lambda=2.5\times10^{-4}$}
          }
\gridline{
          \fig{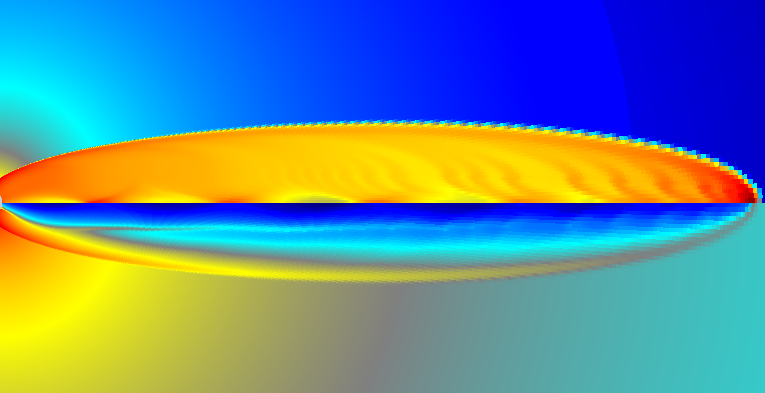}{0.32\textwidth}{(d) $\lambda=2.5\times10^{-3}$}
          \fig{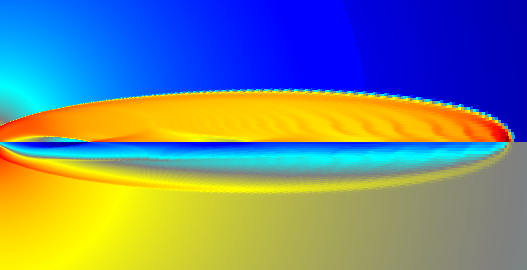}{0.32\textwidth}{(e) $\lambda=2.5\times10^{-2}$}
          \fig{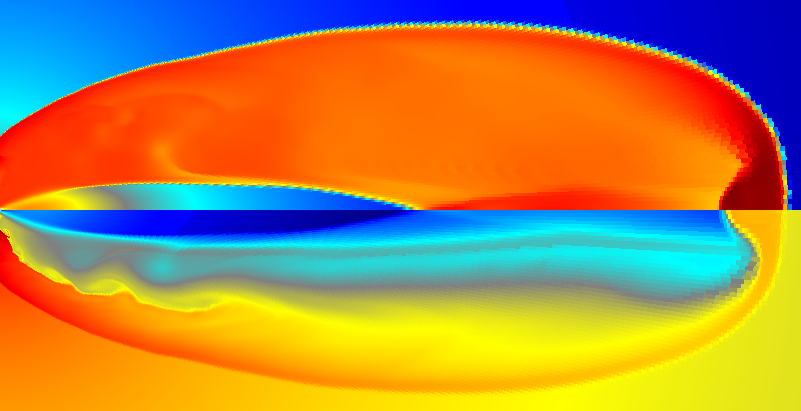}{0.32\textwidth}{(f) $\lambda=2.5\times10^{-1}$}
          }
\gridline{
          \fig{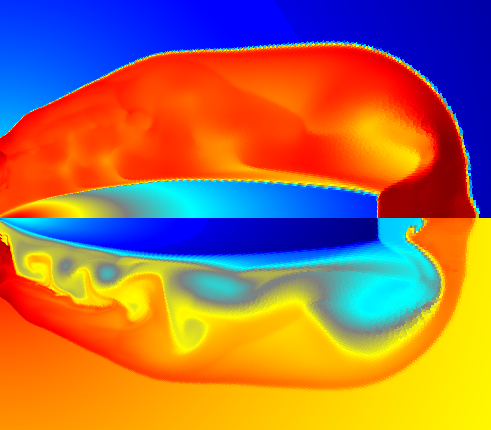}{0.32\textwidth}{(g) $\lambda=2.5\times10^{0}$}
          \fig{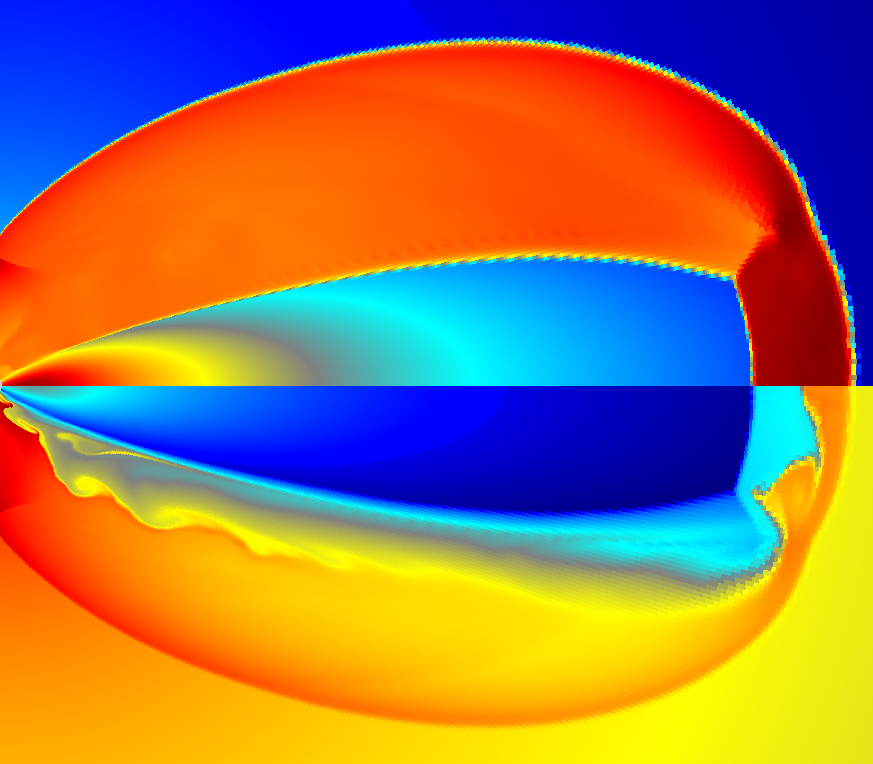}{0.32\textwidth}{(h) $\lambda=2.5\times10^{1}$}
          \fig{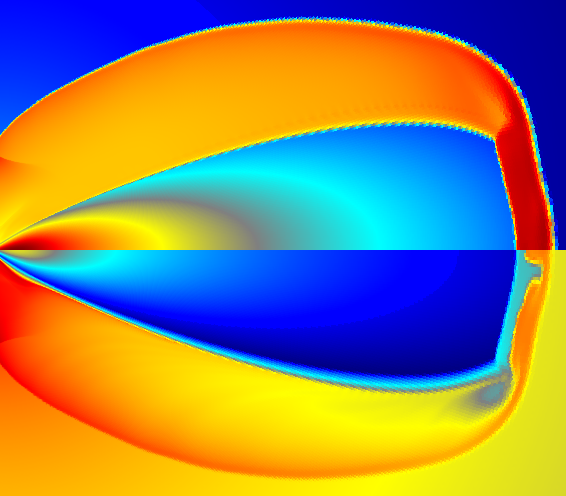}{0.32\textwidth}{(i) $\lambda=2.5\times10^{2}$}
          }
\centering
\includegraphics[width=0.8\textwidth]{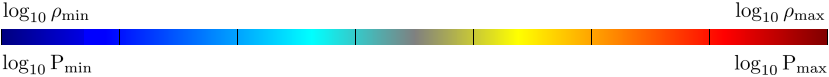}
\caption{Pressure and density maps of self-similar solutions for a single engine opening angle ($\theta_0 = 0.20$ rad) and values of $\lambda$ ranging from $2.5\times10^{-6}$ to $2.5\times10^{2}$. The top and bottom halves of each figure represent pressure and density, respectively. The self-similar solutions range all the way from a non-relativistic forward shock to the ultra-relativistic regime. The colorbar used for plotting the solutions is provided at the bottom. The quantities $\rho_{min}$, $\rho_{max}$, $P_{min}$, and $P_{max}$ denote the minimum and maximum density and the minimum and maximum pressure in the maps, respectively.
\label{fig:zoo}}
\end{figure*}

\section{Diagnostics}   \label{sec:diagnostics}

\subsection{Measurement of dynamical parameters}

We measure the jet head advance speed ($v_h$), the cocoon expansion speed ($v_c$), the final jet opening angle ($\theta_j$), and the cocoon pressure ($P_c$) for each of our numerical solutions. 

The jet head and the cocoon size increase linearly with time. Thus, $v_h$ and $v_c$ can be measured for a solution at a given time by identifying the jet head or the maximum longitudinal expanse of the shock front, and the maximum lateral expanse of the shock front, respectively. The shock front is identified by noting that the maximum pressure drop along any radial direction should occur at the shock front.

The jet opening angle, $\theta_j$, is calculated by measuring the solid angle $\Omega$ subtended by the jet. We calculate $\theta_j$ following \cite{Duffell-Laskar2018} as:

\begin{equation}
\label{eq:theta_j}
    \mathrm{sin}\,\left(\frac{\theta_j}{2}\right) = \frac{ \int (dE/d\Omega) d\Omega }{\sqrt{4\pi \int (dE/d\Omega)^2 d\Omega}}
\end{equation}

We ensure that equation~\ref{eq:theta_j} measures only the angle subtended by the jet by considering the term $dE/d\Omega$ only for those computational zones where the passive scalar, $X = 1$.

The cocoon pressure, $P_c$, is given by the mean pressure of all the cells in our grid corresponding to the cocoon ($0<X<1$), weighted by the cell volume.

\begin{figure*}
\centering
\gridline{\fig{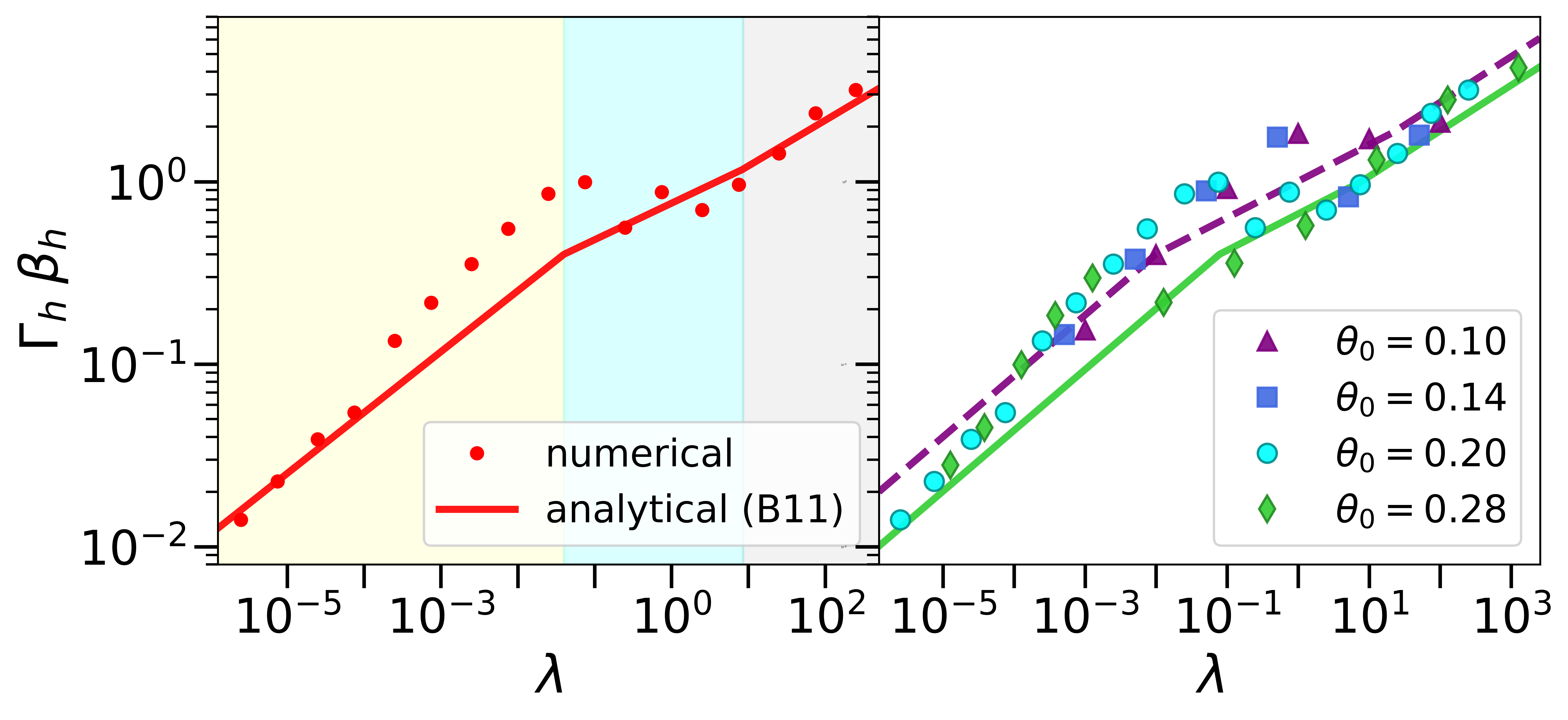}{0.5\textwidth}{(a)}
          \fig{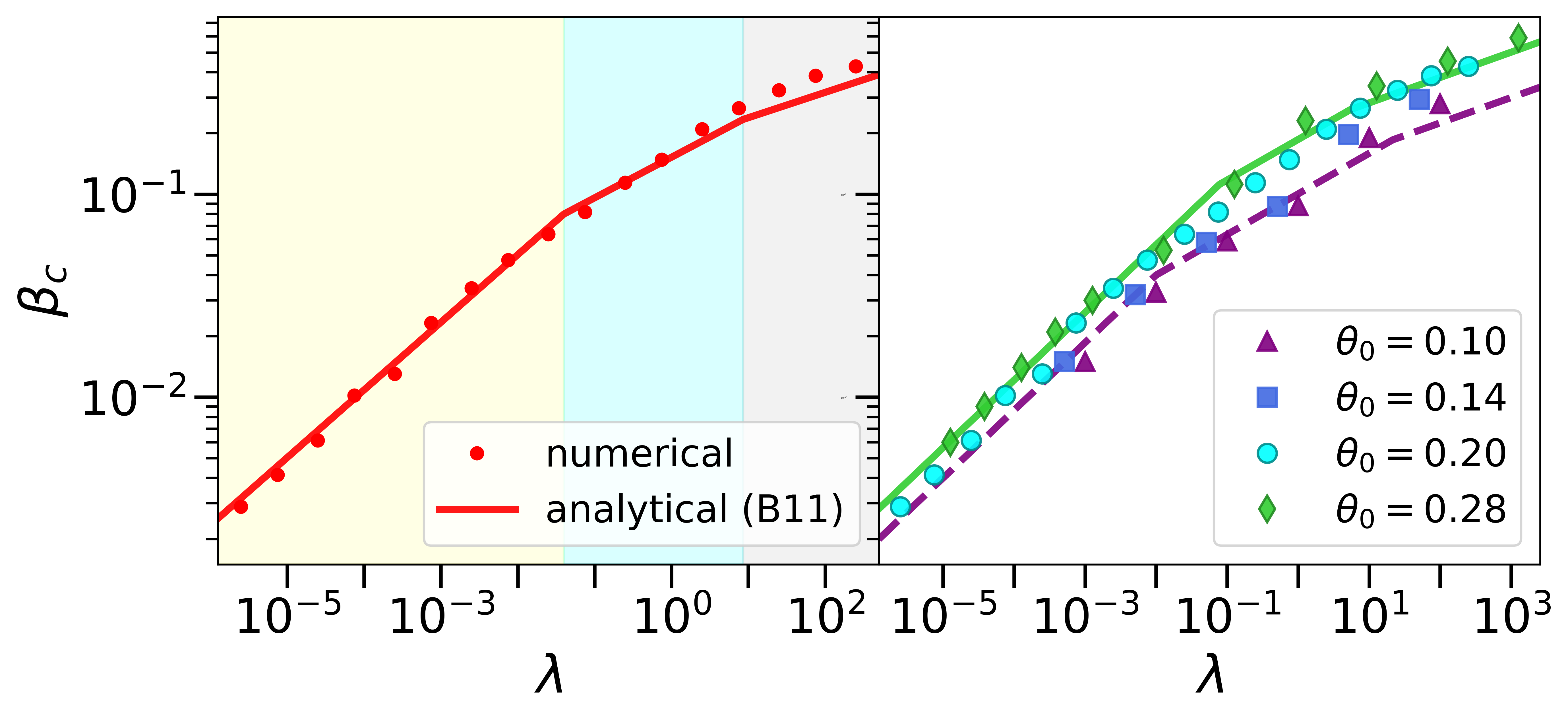}{0.5\textwidth}{(b)}
          }
\gridline{\fig{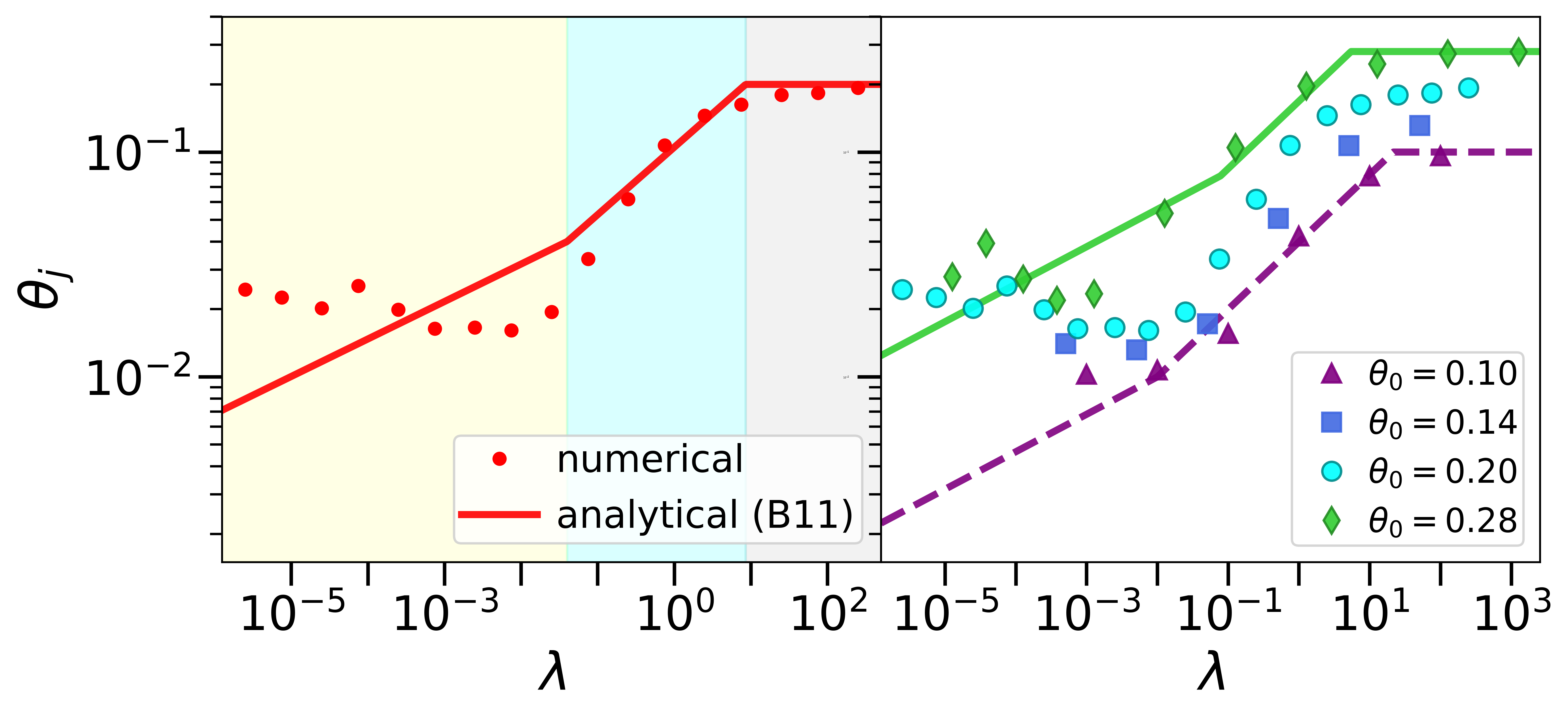}{0.5\textwidth}{(c)}
          \fig{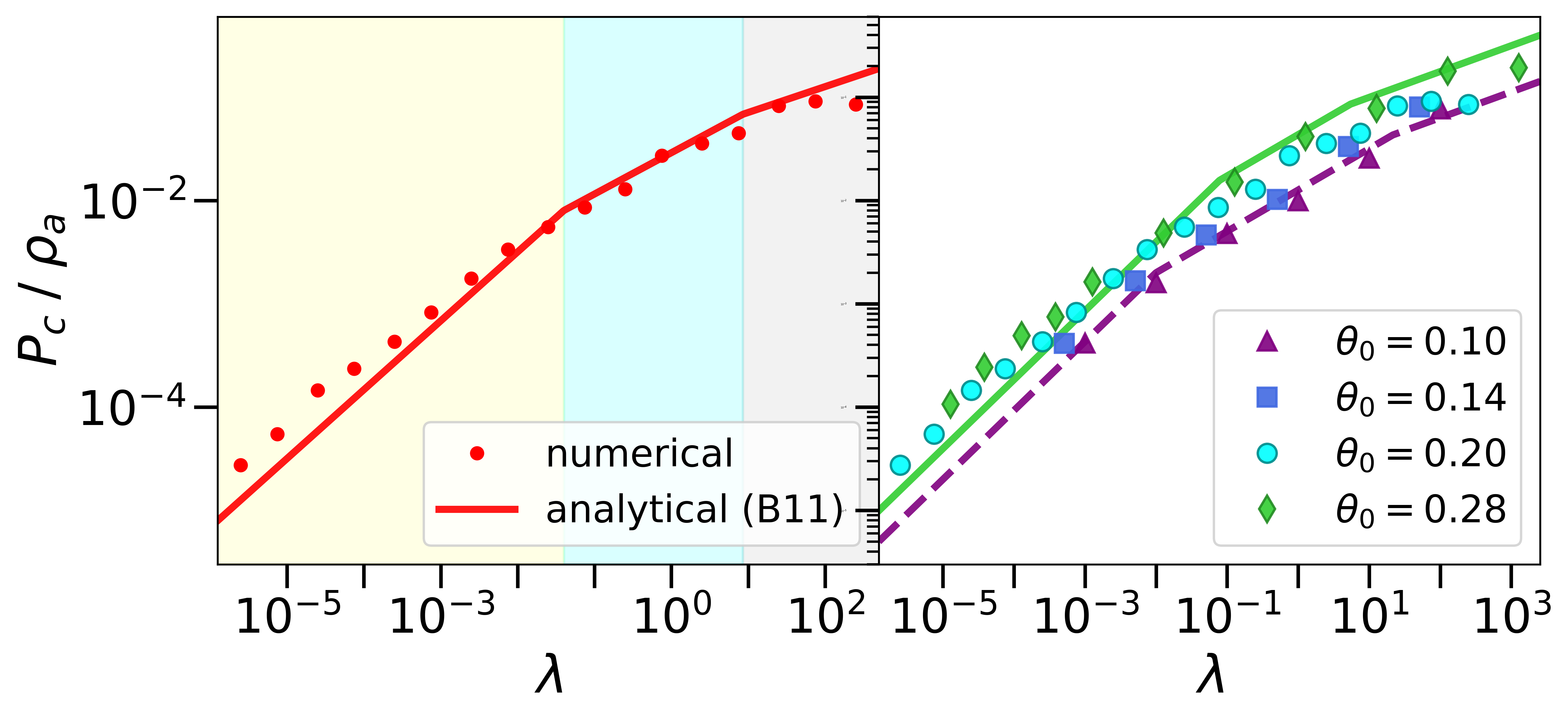}{0.5\textwidth}{(d)}
          }
\caption{Scaling of various dynamic quantities vs $\lambda$. Panels (a)-(d) show the dependence of $\Gamma_h\beta_h$, $\beta_c$, $\theta_j$ and $P_c/\rho_a$ on $\lambda$ respectively. These quantities are the same as in Table~\ref{t:lambdascaling}. The left half of each plot shows the comparison between our result (red dots) and that predicted by \ctB (red solid line) for $\theta_0 = 0.20$. The yellow, blue, and grey shaded regions on plots on the left half indicate the first three dynamical regimes, as in Table \ref{t:lambdascaling}. The right half of each panel show the same scalings for all values of $\theta_0$ used in this work. The purple dashed line and green solid line in the right half of each panel show analytical scalings at $\theta_0=0.10$ and $\theta_0=0.28$ respectively. 
\label{fig:scalings}}
\end{figure*}

\subsection{Emissivity maps} \label{subsec:emissivity}

We also calculate synthetic synchrotron emissivity maps from our results to compare features with observations. For each run, we choose a few viewing angles ranging from 0 to $\pi/2$. It is assumed that the emission from the jet is dominated by synchrotron radiation and that the jets are optically thin. We verify the latter assumption by explicitly calculating optical depth maps for our jets using the method followed by \cite{Westhuizen+2019MNRAS}. The value of magnetic field strength is approximated by requiring a sub-equipartition magnetic field, whose energy density is $2\%$ of the internal energy density in our solutions. We note that this produces field strengths of $\sim0.01-0.1$ mG, similar to values predicted by independent observational studies for jets in M87 \citep{Stawarz+2005ApJ} and Cygnus A \citep{Carilli+1996AandARv}. Our optical depth maps show $\tau<0.01$, thus validating the assumption that our jets should be optically thin.
The emissivity from each computational zone is calculated following a prescription similar to that by \cite{Duffell-Kasen2016arXiv}. However, we assume that the magnetic field is only introduced by the engine and is not significant in the ambient medium. Therefore only material originating from the jet can have nonzero emissivity.  We implement this by using our passive scalar X. The magnetic field strength is assumed to decrease with distance from the central engine as $B(r)\propto r^{-2}$. This emissivity is multiplied by the following factor to account of relativistic beaming \citep{Urry+1995PASP}:

\begin{figure}
\centering
\includegraphics[width=0.45\textwidth]{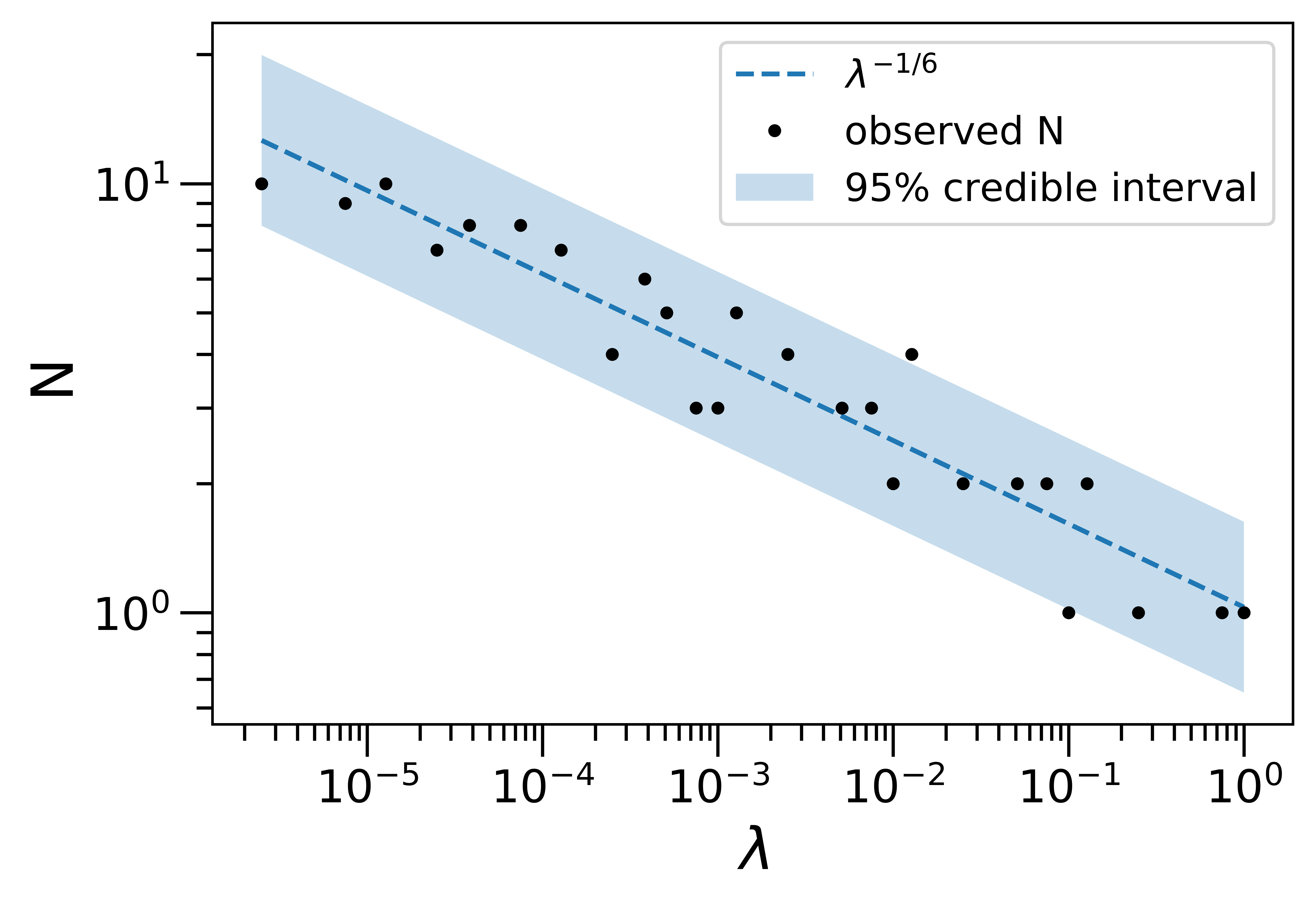}
\caption{Scaling of number of knots vs $\lambda$. The black dots denote the number of knots observed in the emissivity maps. The blue dashed line corresponds to the $N = \lambda^{-1/6}$ curve. The dependence on $\theta_0$ is weak and is accounted for by 95\% credible interval shaded in blue.}
\label{fig:Nscaling}
\end{figure}

\begin{equation}
    D = \left[\frac{1}{\Gamma(1-\vec{\beta}\cdot\hat{n})}\right]^{2+\alpha}
\end{equation}

where $\Gamma$ and $\vec{\beta}$ are the Lorentz factor and velocity of the cell, $\hat{n}$ is the unit vector along the line of sight, and $\alpha$ is the spectral index for synchrotron emission, taken here to be equal to 0.75. We also considered a constant magnetic field and a distance dependence $B(r)\propto r^{-1}$, and found our results are not significantly affected by this choice.

The emissivities are integrated along the line of sight to produce a 2D map of the jet. A dynamic range (in this case, ratio of the maximum pixel intensity to the minimum pixel intensity) is imposed on the maps for them to resemble observations. We choose a dynamic range of $\sim10^4$, as is typical for VLA extended radio emission images of kpc scale jets.

\section{Results} \label{sec:results}

\subsection{General morphological features}

\begin{figure*}
\centering
\gridline{\fig{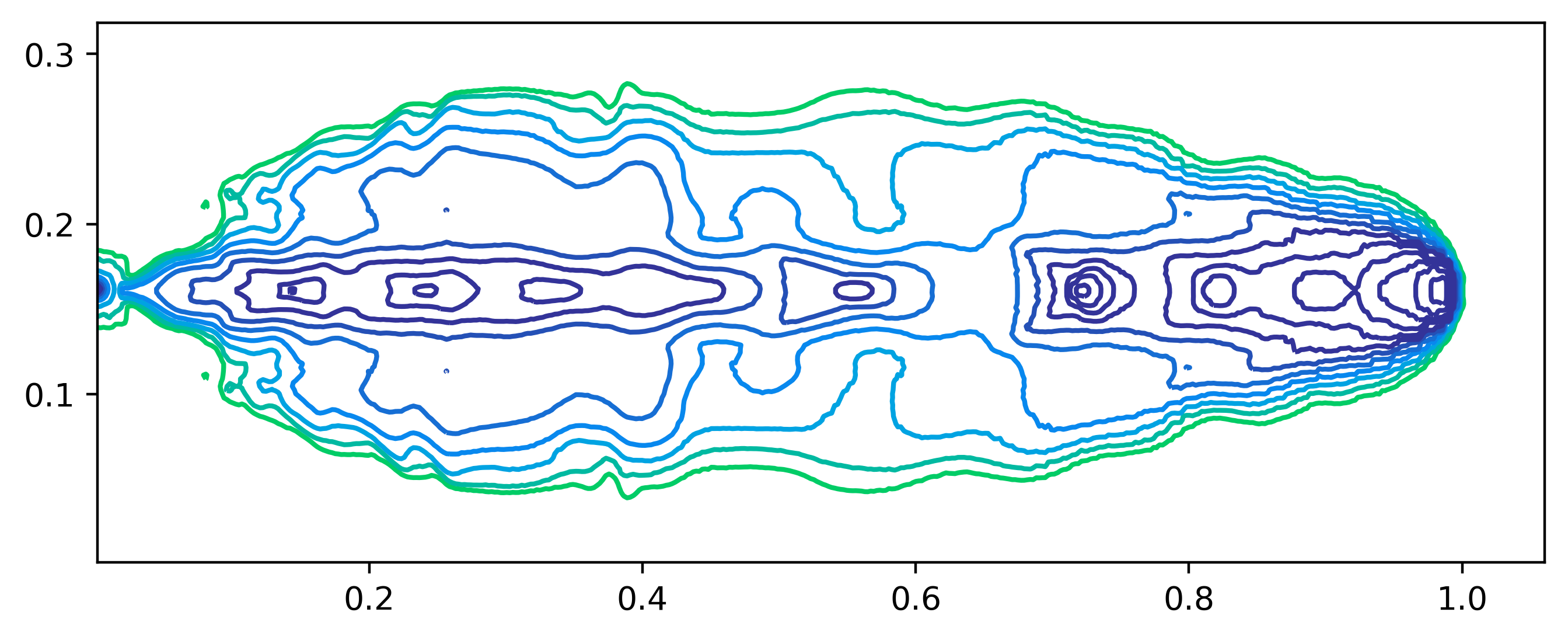}{0.5\textwidth}{(a)}
          \fig{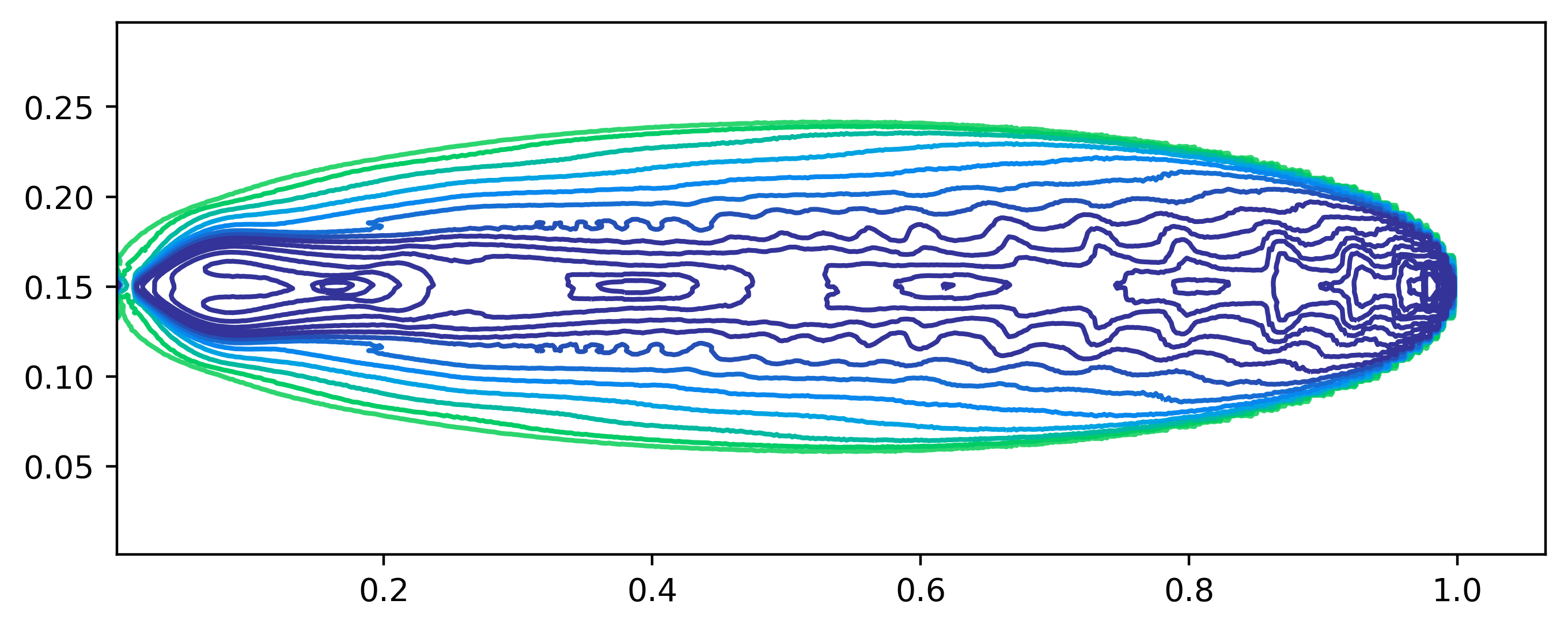}{0.5\textwidth}{(b)}
          }
\gridline{
          \fig{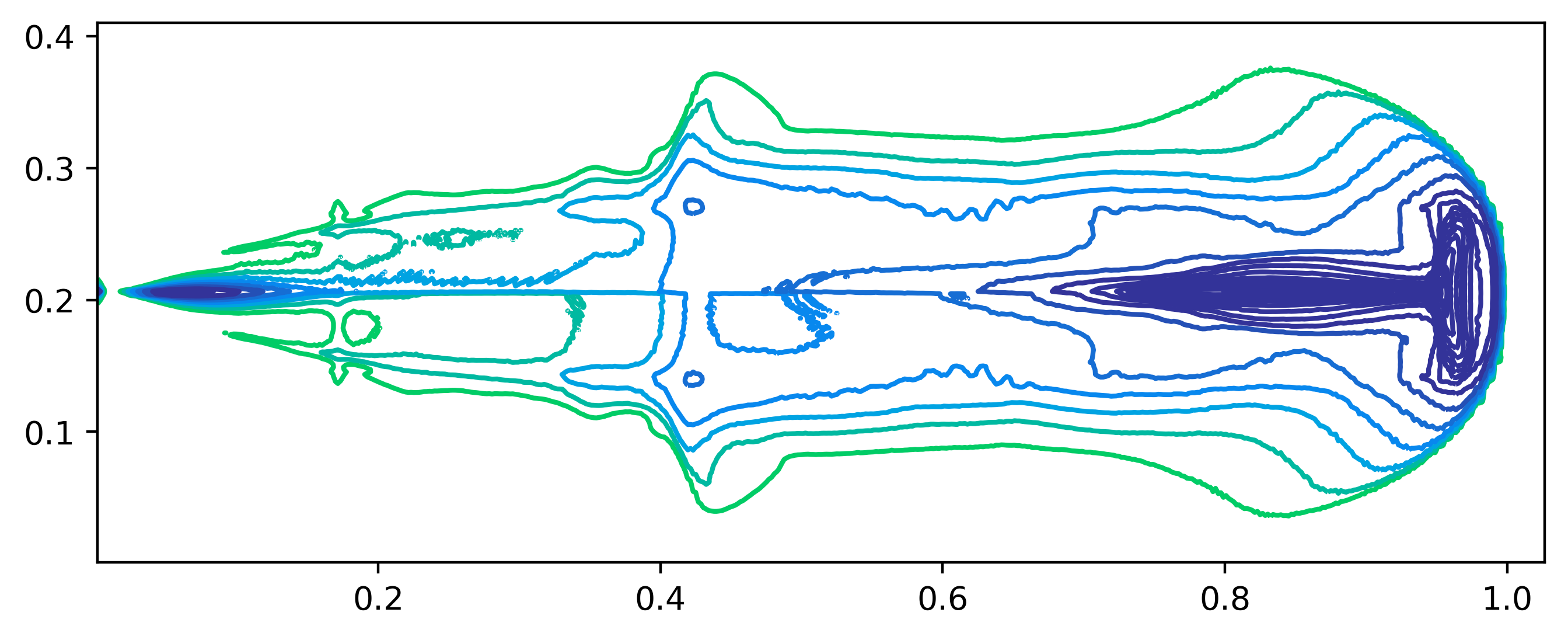}{0.5\textwidth}{(c)}
          \fig{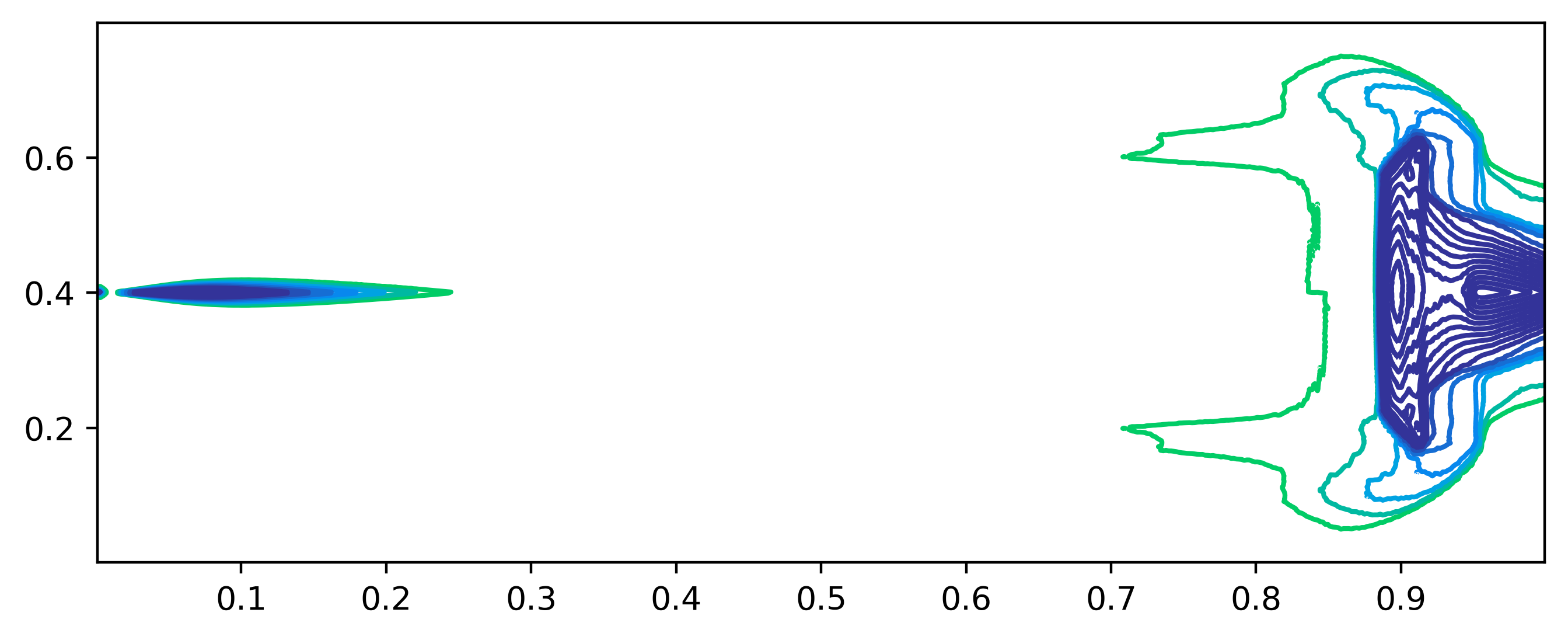}{0.5\textwidth}{(d)}
          }
\includegraphics[width=0.8\textwidth]{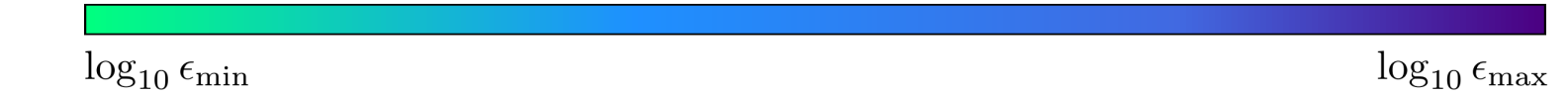}          
\caption{Panels (a)-(d) show the emissivity contour plots for jets with $\theta_0 = 0.20$ and $\lambda$ = $7.5\times10^{-5}$, $7.5\times10^{-3}$, $7.5\times10^{-1}$ and $7.5\times10^{1}$, respectively. The viewing angle is $90^{\circ}$ in each case. Multiple recollimation shocks are seen in panels (a) and (b), while panels (c) and (d) exhibit one or no recollimation shocks. Each contour level is brighter than the preceding contour level by a factor of $\sqrt{2}$. The brightest contour level is a factor of $10^4$ times brighter than the base contour level for all maps, in accordance with the typical dynamic range of VLA observations. The numbers on the axes denote distances along them in units such that $\mathrm{z_h} = 1$. The quantities $\epsilon_{min}$ and $\epsilon_{max}$ denote the minimum and maximum emissivity per pixel, respectively.
\label{fig:emissivity_maps}}
\end{figure*}

Our fiducial runs, assuming an inverse square power-law atmosphere, show similar qualitative features as predicted by \ctB. Fig.~\ref{fig:example} shows an example plot for a self-similar jet with a low value of $\lambda$. An overpressured cocoon forms around the jet. The jet is seen to be well-collimated, with many recollimation shocks along the axis. The variation of the nature of the jet-cocoon system with increasing $\lambda$ is seen in Fig.~\ref{fig:zoo}. Here we show a subset of our runs with $\theta_0=0.20$ and $k=2$ for a range of engine luminosities. The cocoon seems to be a common feature in all the jets. However, the degree to which jets are collimated decreases with $\lambda$ as the jet head grows wider and the first recollimation shock moves farther from the engine. This trend has been also been reported by previous numerical RHD computations, e.g., \cite{Yates+2018MNRAS}. Almost all the low $\lambda$ jets exhibit multiple recollimation shocks.

\subsection{Scalings of dynamical parameters}

Fig.~\ref{fig:scalings} shows the scalings for jet head velocity, cocoon transverse expansion velocity, jet opening angle, and the ratio of the cocoon pressure to the ambient density at the jet head. The left half of the panels show the scalings with $\lambda$ for $\theta_0=0.2$ radians. They also depict the dynamical regimes from Table~\ref{t:lambdascaling}, which are marked with three different colors (yellow, blue, and gray). It is seen that the transverse expansion velocity of the cocoon ($\beta_c$) scales almost exactly as predicted by \ctB. In contrast, our observed scaling of the jet opening angle ($\theta_j$) appears to deviate from that predicted by \ctB for lower values of $\lambda$. We obtain higher values for $\theta_j$ than is expected by \ctB and $\theta_j$ seems to flatten to a minimum value. It can also be noticed that our measured values of the jet head's four-velocity $\Gamma_h \beta_h$ may indicate a slight deviation from the prediction by \ctB in the intermediate ``Relativistic Head" regime. We further discuss these discrepancies in Section~\ref{subsec:previous_works}.

\subsection{Occurrence of bright spots in the jet}

Bright, non-terminal spots (called knots) are routinely observed in extended emission from AGN jets, at both radio and X-ray wavelengths. Some of the prominent examples are M87 \citep{Marshall2002ApJ}, 3C 273 \citep{Marchenko2017ApJ} and OJ287 \citep{Marscher2011ApJ}. These knots are often attributed to site of particle acceleration in strong shocks in the outflow. These shocks could be formed hydrodynamically by recollimation of the jet by the ambient medium \citep{Komissarov+1998MNRAS}, or from inhomogeneity introduced in the jet by intermittent engine activity \citep{Stawarz2004ApJ}, or by instabilities in the outflow \citep{Micono+1999A&A}. Other alternative theories for knot formation in jets include non-uniform Doppler boosting and sudden large scale expansion in the outflow \citep{Harris2010IJMPD}.

We observe multiple bright spots in some of our synthetic emissivity maps coincident with locations where the recollimation shock(s) converge to the jet axis, resulting in a local pressure maximum. The number of these bright spots or knots can be obtained via visual inspection of the emissivity maps, or by calculating the number of local maxima of pressure along the jet axis. We find both methods to be consistent. The number of knots is found to decrease with increasing $\lambda$, obeying a weak scaling of $N \approx \lambda^{-1/6}$. The effect of varying the injection angle $\theta_0$ is negligible, provided $\lambda$ is held constant. This relationship therefore provides an independent consistency check on the estimated value of $\lambda$ for an AGN jet, using the number of knots observed. A plot of the number of knots in a jet against the jet $\lambda$ value is shown in Fig. \ref{fig:Nscaling}, along with our predicted scaling rule for the same and a 95\% confidence interval.

\subsection{Shapes of AGN jets}

\begin{figure*}
\centering
\includegraphics[width=0.85\textwidth]{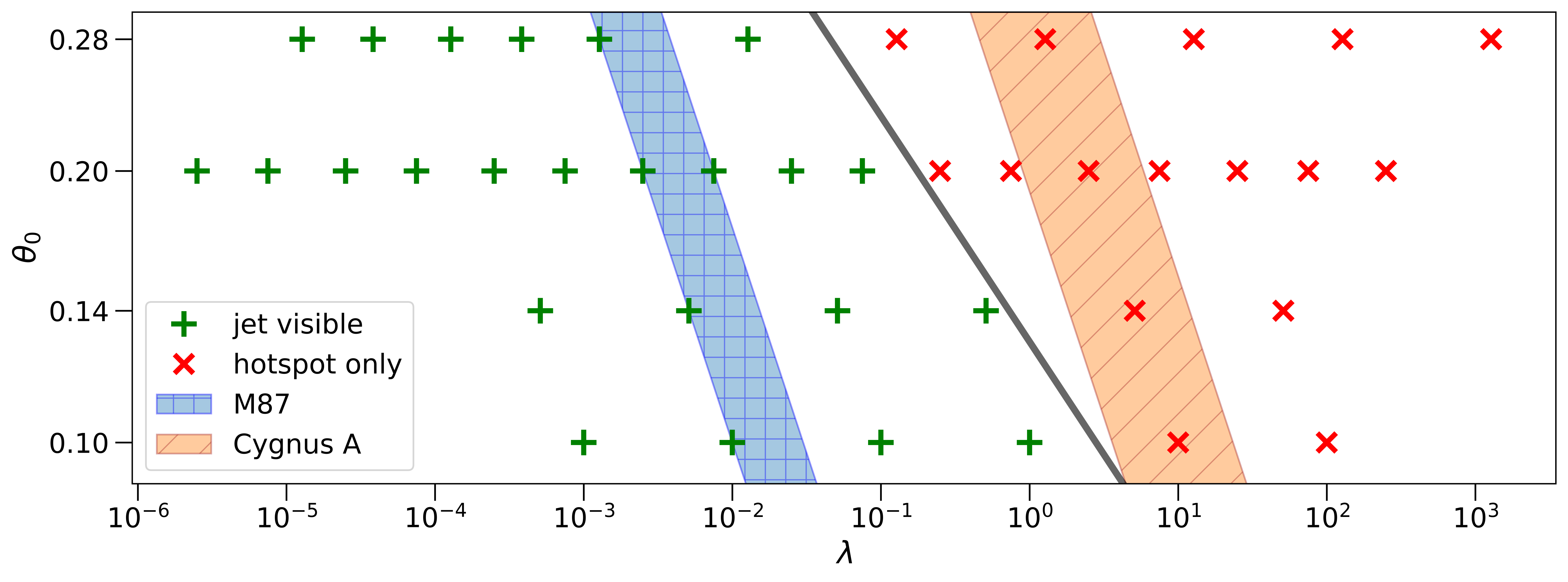}
\caption{Overall morphology of all self-similar solutions. Green plus signs represent a completely visible jet morphology, as seen in panels (a) and (b) in Fig.~\ref{fig:emissivity_maps}. Red cross signs indicate morphology dominated by a prominent hotspot with a somewhat visible cocoon but no visible jet. Panel (c) and (d) in Fig.~\ref{fig:emissivity_maps} are examples of such a morphology. The thick black line represents $\lambda_{\mathrm{crit}}(\theta_0)$ (Inequality \ref{eq:FRC_lambda}), the criterion that distinguishes FRI jets from FRII jets. The blue square-hatched and orange cross-hatched regions represent the approximate location that M87 and Cygnus A (respectively) are most likely to occupy on this diagram. }
\label{fig:morphology_distinction}
\end{figure*}

A subset of the synthetic emissivity maps for a range of $\lambda$ is shown in Fig.~\ref{fig:emissivity_maps}. It is seen they exhibit two distinct types of morphology: the low-$\lambda$ systems show a distinct jet structure with one or more knots, while only the bright jet head is visible in high-$\lambda$ systems. This is strongly reminiscent of the Fanaroff-Riley dichotomy. The original suggestion was that center-brightened or FRI galaxies have lower radio luminosities, while the edge-brightened or FRII galaxies have higher radio luminosities. It has been since suggested that the FRI/FRII morphological divide may instead be related to both the jet power and the environmental density (see \cite{Hardcastle+2020NewAR} and references therein). A jet with a given luminosity may remain relativistic and terminate in a bright hotspot in a poor environment, but may decelerate due to entrainment of ISM in a denser environment, resulting in an FRI structure. Our result strongly corroborates with this hypothesis. We find that such a morphological dichotomy should also depend on the injection angle $\theta_0$. This is depicted in Fig.~\ref{fig:morphology_distinction}, where we plot the two different types of morphology shown by our results on a $\theta_0$ vs $\lambda$ diagram. Fig.~\ref{fig:morphology_distinction} also shows the approximate regions most likely to be occupied by the AGN jets in M87 and Cygnus A based on estimates of their $\lambda$ values. This is described in detail in Section~\ref{subsec:observations}.

The demarcation observed in Fig.~\ref{fig:morphology_distinction} can be approximately expressed in terms of $\lambda$ and $\theta_0$, or equivalently in terms of $\Tilde{L}$ and $\theta_j$ as follows:

\begin{equation}
    \lambda_{\mathrm{crit}} \theta_0^4 \lesssim 3\times10^{-4} \label{eq:FRC_lambda}
\end{equation}    
    
or,

\begin{equation}
    \Tilde{L}_{\mathrm{crit}}^{1/4} \theta_j^3 \lesssim 2\times10^{-3} \label{eq:FRC_ltilde}
\end{equation}

In other words, AGN jets should exhibit an FRI morphology if they satisfy inequalities~\ref{eq:FRC_lambda} or~\ref{eq:FRC_ltilde}, and an FRII morphology otherwise. The region of the parameter space where the FRI/II transition happens corresponds to a collimated jet with a relativistic head. Therefore we assume relations between $\lambda$, $\Tilde{L}$, $\theta_0$ and $\theta_j$ appropriate to that regime while deriving equation~\ref{eq:FRC_ltilde} from equation~\ref{eq:FRC_lambda}.

\subsection{Environments that Break Self-Similarity} \label{sec:k1}

AGN are usually observed in host environments with density profiles shallower than $\rho(r)\propto r^{-2}$ ($k\sim1\mbox{--}1.5$, as explained in Section~\ref{subsec:IC}). It can be seen from equation \ref{eq:lambda} that the value of $\lambda$ should decrease with time for a jet advancing through such an environment. Since the jet head velocity and the jet opening angle depend on $\lambda$, the jet decelerates and changes shape as it advances. This implies AGN jets may not evolve self-similarly.

We investigate whether the scalings and general morphological trends obtained from our self-similar models hold for jets in observed AGN host environments. This is done using a single run, corresponding to the last row in Table \ref{t:engine}, where $k=1$ and $\theta_0=0.2$. The jet starts out with a high $\lambda_{\mathrm{eff}}$ value and is evolved for nine orders of magnitude in time. $\lambda_{\mathrm{eff}}$ is inversely proportional to the jet length $z_h$ in this case according to equation \ref{eq:lambda} since $\rho_a(z_h)\propto z_h^{-1}$. Thus $\lambda_{\mathrm{eff}}$ decreases as the jet grows in size. The duration of evolution is long enough to let $\lambda_{\mathrm{eff}}$ cover a range of values comparable to the range of $\lambda$ for the self-similar models (with $k=2$ and $\theta_0=0.2$). 

\begin{figure}
\centering
\includegraphics[width=0.48\textwidth]{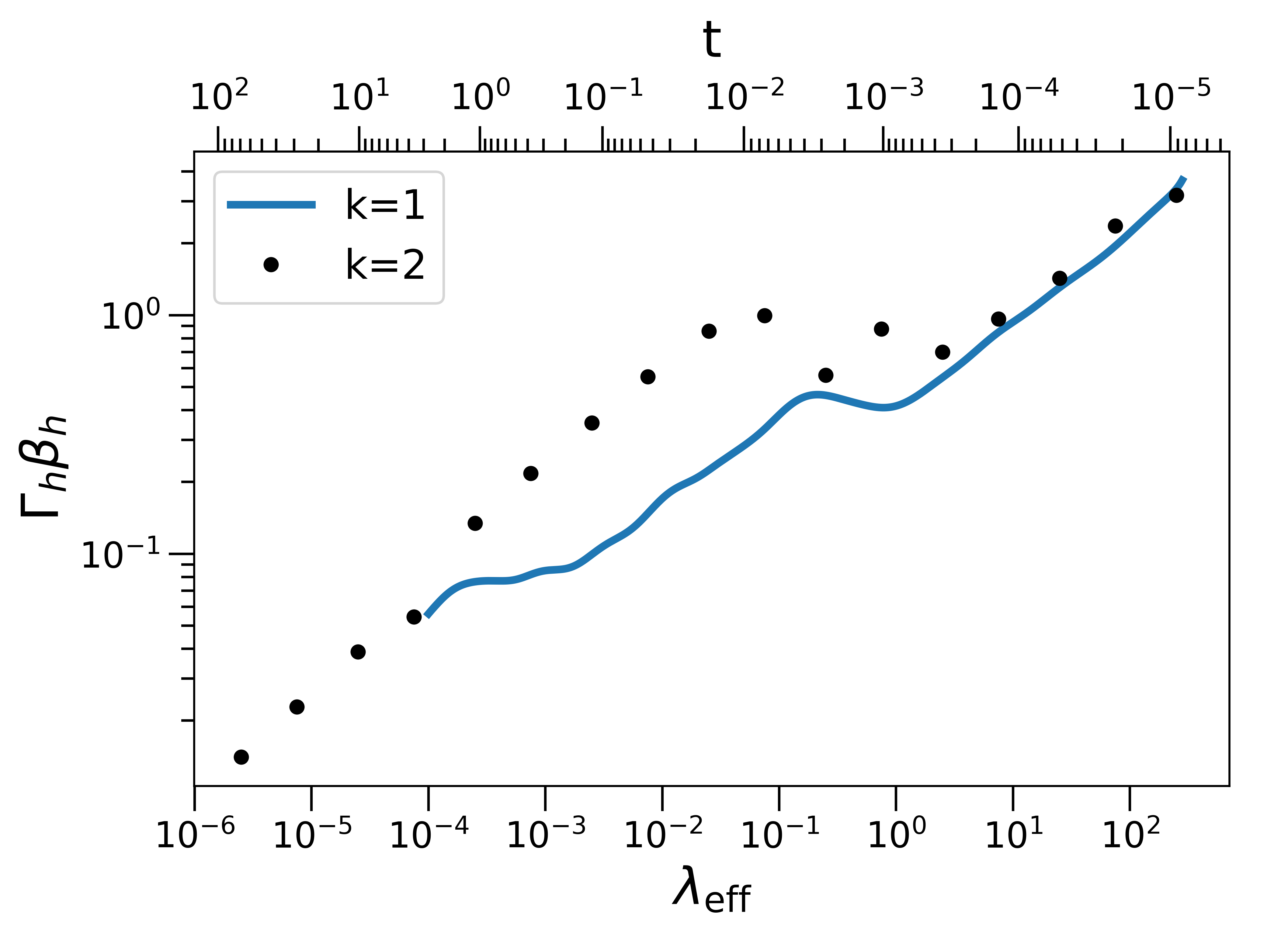}
\caption{Jet head proper velocity ($\Gamma_h\beta_h$) as a function of $\lambda_{\mathrm{eff}}$ for the jet expanding against the k=1 density profile, denoted by the blue solid line. The black dots show $\Gamma_h\beta_h$ for the self-similar (k=2) jets as a function of $\lambda$. The upper x-axis values show the time (in code units) it took the k=1 jet from starting to attain the corresponding value of $\lambda_{\mathrm{eff}}$ on the lower x-axis. }
\label{fig:k1_dynamics}
\end{figure}

The jet head position is measured at closely spaced instants of time. The numerical derivative of the head position with respect to time provides us the head velocity as a function of time, or equivalently as a function of $\lambda$. Fig.~\ref{fig:k1_dynamics} shows a comparison between the dependences of the proper head velocity on $\lambda$ for the self-similar jet (k=2) and the decelerating jet (k=1). The self-similar results appear to be applicable to non-self-similar jets (k=1). This suggests that the solution roughly transitions from high-$\lambda$ to low-$\lambda$ solutions consistent with our self-similar results, even though the jet does not evolve self-similarly in this case. This shows that the dynamics of a jet does not significantly depend on how $\lambda$ changes with time. Thus, the instantaneous value of $\lambda$ for a jet is enough to characterize its dynamical properties. Interestingly, the time evolution of $\lambda$, as indicated by the upper x-axis in Fig.~\ref{fig:k1_dynamics} suggests that FRII sources should transform into FRI sources given sufficient time so that $\lambda$ falls below unity. A similar conclusion was reached by \cite{Li+2018ApJ}. We use Inequality~\ref{eq:FRC_lambda} or~\ref{eq:FRC_ltilde} assuming parameters relevant to Cygnus A to deduce what length such a jet has to grow to before such a transition occurs. We find that even though Cygnus A barely satisfies our FRII criterion, it has to grow from $\sim70$ kpc to 700 kpc before we expect it to show FRI morphologies. Hence, we do not expect to observe transition of known FRII sources to FRI.

\section{Discussion}  \label{sec:discussion}

\subsection{Comparison with observations} \label{subsec:observations}

We obtain a rough estimate of $\lambda$ for two archetypal nearby FRI and FRII jets, in the radio galaxies M87 and Cygnus A, respectively to see where they lie on Fig. \ref{fig:morphology_distinction}. Equation \ref{eq:lambda} is used to express $\lambda$ as follows:

\begin{equation}
\label{eq:M87-lambda}
    \begin{split}
        \lambda \approx 2\times10^{-2} \times \left(\frac{L}{10^{43}\,erg\,s^{-1}}\right) \times \left(\frac{\rho}{10^{-28}g\,cm^{-3}}\right)^{-1} \\ \times \left(\frac{z_h}{1.5\,kpc}\right)^{-2} \times \left(\frac{\theta_0}{0.1\,rad}\right)^{-2}
    \end{split}
\end{equation}

The value of $\lambda$ for the jet in M87 is calculated to be approximately equal to $\sim2\times10^{-2}$, using observations for the kiloparsec scale jet \citep{Biretta+1995ApJ}, density in the neighborhood of the M87 nucleus \citep{Russell+2015MNRAS}, and jet power estimates \citep{Stawarz+2006MNRAS,Russell+2013MNRAS}. Cygnus A, on the other hand, has a $\lambda$ value of $\sim 9$, calculated from a jet power estimate $\sim 10^{46}\,\mathrm{erg\, s^{-1}}$, jet head density $\sim 10^{-31}\,\mathrm{g\, cm^{-3}}$ \citep[][and references therein]{Godfrey+2013ApJ,Snios+2018ApJ} and a core to hotspot distance of $\sim 70\,\mathrm{kpc}$ \citep{Carilli+1996AandARv}.

The major sources of error in our estimates for $\lambda$ are from uncertainty in jet power estimates and our ignorance about the engine opening angle. We expect each of these to contribute an uncertainty of a factor of few. For example, the jet power values used in this work have been estimated by dividing the amount of mechanical work done by the jet to inflate the radio lobes observed around it by the buoyancy timescale of the lobes \citep{Birzan+2004ApJ}. This estimate neglects radiated power and suffers from time-averaging effects as well as uncertainties in lobe volume or pressure calculation. In case of M87, additional uncertainty of at most a factor of 2 comes from the position of the jet head. We assume it to be co-spatial with the outermost knot (knot C). It's unclear if the jet corresponding to our models extends beyond this region, as seen in high resolution radio images of M87 \citep[e.g., Fig. 1 of][]{Biretta+1995ApJ}. Nonetheless, our results therefore correctly classify the Fanaroff-Riley morphology of the jets in M87 and Cygnus A. This is demonstrated in Fig.~\ref{fig:morphology_distinction}, where we plot the approximate locations of M87 and Cygnus A. The prediction of our classification criterion (Equation~\ref{eq:FRC_lambda}) holds even allowing for the error in jet power and engine opening angle estimates. We calculate whether M87 and Cygnus A satisfy inequality~\ref{eq:FRC_ltilde} as a sanity check, since $\theta_j$ and therefore $\Tilde{L}$ are usually easier to estimate from observations and find our predictions to be consistent. 

Our prediction for the number of knots in the jet of M87, from Fig.~\ref{fig:Nscaling} turns out to be $\sim 4$. This is somewhat less than the 6 or 7 resolved knots seen in the kpc scale jet, but may be attributed to the uncertainty in our knowledge of $\lambda$. Moreover, the $\lambda$ value for the jet in Cygnus A is very close to the threshold value, which indicates it may exhibit multiple recollimation shocks or knots characteristic of FRI jets. That is exactly what is seen in the inner kpc scale or parsec scale images of Cygnus A \citep[figure 4 of][]{Carilli+1996AandARv}.

\subsection{Comparison with previous works} \label{subsec:previous_works}

Our scaling relations shown in Fig~\ref{fig:scalings} indicate that $\theta_j$ almost stays constant for low values of $\lambda$ as opposed to the $\theta_j \propto \lambda^{1/6}$ scaling predicted by \ctB for $\lambda<\theta_0^2$. Therefore jet collimation appears to be increasingly less effective compared to the analytical expectation as $\lambda$ or $\Tilde{L}$ decreases. The opening angle that the system converges to at low $\lambda$ is seen to be a function of $\theta_0$ and is consistent with the scaling $\theta_0^{1/2}$. We investigated the possibility that this disagreement arose simply due to our inability to resolve the inner zones close to the recollimation shock. \cite{Harrison+2018MNRAS} formulate a criterion for their numerical solutions to ensure the nozzle size is adequate for capturing jet dynamics accurately. They find that the calculations are consistent as long as the first recollimation shock radius and height are much greater than the injection radius and height, respectively. We found all our solutions satisfy this criterion. Additionally, we repeated one of our calculations at a low value of $\lambda$ ($2.5\times10^{-4}$) with an inner boundary and an injection radius both smaller by a factor of 2 than the rest of the runs. The same value was found for $\theta_j$, suggesting the jet is indeed less collimated than expected.  One possibility is that in this very low-$\lambda$ regime the solution might revert to entirely non-relativistic scalings (even though the jet core is still relativistic), which would imply a jet and cocoon morphology completely independent of $\lambda$.  So far this idea is speculative, but may be worth investigating in a follow-up study.

The jet head's four-velocity $\Gamma_h \beta_h$ has a very shallow dependence on $\lambda$ in the intermediate regime. \ctB predict a very weak scaling in this regime, and our points have sufficient scatter that they are still consistent with \ctB, but they are equally consistent with no dependence on $\lambda$ at all in this regime, only depending on $\theta_0$.  It would be interesting to investigate this in more detail in a future study, as very minor changes in the scaling of $\theta_j$ with $\lambda$ would be sufficient to create such a flat scaling.

The recollimation shock ``trains" seen in our solutions have been reported by previous numerical studies of relativistic jets, such as \cite{Perucho+2007MNRAS} and \cite{Saxton2010MNRAS}. The latter show that regularly spaced knots are formed in a supersonic jet moving against constant density ambient medium. However, they predict that this spacing should decrease if the ambient density falls off with distance. We obtain roughly constant knot spacing in our relativistic jet models for a given engine luminosity and opening angle irrespective of the ambient density profile, as is seen in observations \citep{Godfrey2012ApJ}. We also obtain the explicit scaling of the number of knots with the system parameters.

Our solutions exhibit the major dynamical and morphological features seen in long term evolution using numerical RHD computations of both FRI \citep{Perucho+2007MNRAS} and FRII \citep{Perucho+2019MNRAS,Perucho+2022MNRAS} jets. In particular, the zoo of our jet models show a morphological dichotomy that primarily depends on $\lambda$ and weakly on $\theta_0$. The emissivity map of the high $\lambda$ jets is dominated by the jet head, while the low $\lambda$ jets exhibit the complete jet structure, often accompanied by several bright knots on the jet axis. This is consistent with the Fanaroff-Riley dichotomy, which has been suggested to be governed primarily by the jet power in some numerical studies \citep{Li+2018ApJ,Seo+2021ApJ}. We argue that the deciding parameter is $\lambda$, or the ratio of jet power to ambient density. \cite{Li+2018ApJ} also examine the impact of jet speed and the jet to surroundings density contrast and show that for a given jet power, jets with low density in the jet core ($\rho_j$) and/or low Lorentz factor ($\Gamma_j$) should exhibit FRI morphology, while high density and/or high Lorentz factor jets should form FRII morphologies. This is in rough agreement with our results, since $\rho_j$ and $\Gamma_j$ are directly proportional to $\Tilde{L}$ or equivalently $\lambda$, according to equation \ref{eq:Ltilde}.

\subsection{Scope for future work} \label{subsec:future_work}

It would be interesting to study some important aspects of AGN jets in the future which have not been addressed in this work. We don't take into account the role of magnetic fields in collimating the jet. 
We consider the well-studied jet in M87 to determine if the magnetic field is dynamically important at kpc scales. \cite{Stawarz+2005ApJ} show using $\gamma$-ray observations that flux from the brightest knot indicates a magnetic field strength of $30\mbox{--}100\,\mathrm{\mu G}$. As mentioned in Section~\ref{subsec:emissivity}, our solutions require an equipartition parameter of ~0.02 to reproduce this value. This implies magnetic fields are not dynamically important for our solutions, which apply to kpc scale jets.
But there may be a deviation from our predictions in Section~\ref{sec:k1} due to magnetic fields for smaller scale jets. Additionally, magnetic fields may play an important role in producing the Fanaroff-Riley dichotomy, as has been suggested by \cite{Tchekhovskoy+2016MNRAS}.

We do not solve for 3D jets. Some numerical studies suggest that there may be some differences in the jet head and the jet spine structure for 2D and 3D models. \cite{Rossi+2008A&A} and \cite{Matsumoto+2021MNRAS} observe that mixing between the jet and the cocoon disrupts the jet spine structure for some of their 3D jet models, especially upstream of the first recollimation shock. Thus these models do not show the recollimation shock train structure that we'd expect from our 2D models. Similar differences pertaining to jet-cocoon mixing and jet head structure are seen by \cite{Harrison+2018MNRAS} who compare 2D and 3D jet calculations in the context of calibrating the analytical model by B11. However, they find that the general morphology of the jet, the cocoon and recollimation shocks remain similar for 2D and 3D jet models. Some other numerical studies also report multiple recollimation shocks in 3D hydrodynamic jet models, e.g., \cite{Zhang+2004ApJ} in the context of a jet propagating through a massive star, \cite{Mukherjee+2018MNRAS} in the context of AGN jets propagating through a turbulent galactic disk and the galactic halo, and \cite{Wang+2008ApJS} for relativistic jets in general. \cite{Rossi+2008A&A} note that formation of multiple recollimation shocks seems more plausible for their relativistic jet models as opposed to non-relativistic ones. 

In general, recollimation shocks are seen to be weaker in 3D than in 2D studies, and the conical structure seen at the jet head seems to be purely a 2D artifact. Thus it is unclear if the recollimation shock train seen in our 2D models would remain the same in 3D simulations. It is also unclear how the emissivity map of strong jets with bright heads would look like for 3D solutions, which haven't been calculated in this work.

Lastly, the dynamics and morphology of AGN jets may depart from our constant luminosity jet models if engine variability is taken into account, which is observed for AGN on time scales ranging from minutes to years.

\section{Conclusion}  \label{sec:conclusion}

We have obtained self-similar models of relativistic jets expanding against ambient medium for a wide range of system parameters. We have compared these results to the long-term evolution of a jet in a cluster-center like ambient medium (with a density profile $\rho(r) \propto r^{-1}$), where the jet no longer expands self-similarly, and found our self-similar results to remain applicable for suitable definitions of the parameters $\lambda$ and $\theta_0$. Thus, the dynamics of the system should only depend upon the instantaneous values of $\lambda$ and $\theta_0$ and not on the history of evolution as long as the density profile can be modeled as a modest power-law. This eliminates the need for evolving the jet over many orders of magnitude in time, allowing us to infer dynamical parameters of AGN jets by measuring the instantaneous value of $\lambda$ through observations.

The model developed by \ctB has been found to hold for relativistic jets with some minor modifications to the jet opening angle scaling. In particular, the jet opening angle seems to not decrease with $\lambda$ for non-relativistic jets (equivalently, for low values of $\lambda$) but attains a constant value at a low enough $\lambda$. We do not yet have an explanation for this behavior at low $\lambda$, but it suggests an upper limit to the degree of collimation the jet can experience via the surrounding cocoon.

We obtain a criterion for the Fanaroff-Riley classification of AGN jets that is dependent not only on the jet power but also its opening angle and surrounding medium. FRI jets, in addition to being center-brightened, exhibit multiple bright spots called knots along their axis. Our calculations recover this result as a purely hydrodynamical effect, without needing to include magnetic fields. We find that the number of knots in FRI jets scales with $\lambda$ ($N(\lambda)\propto\lambda^{-1/6}$). This provides a consistency check on the idea these knots are recollimation shocks. M87 and Cygnus A seem to be in rough agreement with this check, although agreement in the case of M87 would require the jet power to be on the low end of observationally inferred values.

\acknowledgments

We acknowledge M. Lister, E. Nakar, and K. Blundell for helpful comments. We thank the anonymous referee for useful suggestions. High-resolution calculations were provided in part by the resource Stampede2 through the local allocation awards \#TG-AST190029 and \#TG-AST190033 in the Extreme Science and Engineering Discovery Environment (XSEDE) supported by National Science Foundation grant number ACI-1548562 \citep{xsede}. We also acknowledge access to the resource Bell at the Rosen Center for Advanced Computing (RCAC) of Purdue University \citep{McCartney2014}.

\bibliographystyle{apj} 
\typeout{}
\bibliography{ms}

\end{document}